\documentclass[aps,prl,twocolumn,groupedaddress]{revtex4}
\usepackage{amsmath}
\usepackage{amssymb}
\usepackage{graphicx}

\begin{document}

\title{Conductivity of graphene: How to distinguish between samples
with short and long range scatterers}

\author{Maxim Trushin and John Schliemann}

\affiliation{Institute for Theoretical Physics, University of Regensburg,
D-93040 Regensburg, Germany}

\begin{abstract}
Applying a quasiclassical equation to
carriers in graphene we found a way how to distinguish
between samples with the domination of short and long
range scatterers from the conductivity measurements.
The model proposed explains recent transport experiments
with chemically doped as well as suspended graphene.
\end{abstract}

\maketitle

\section{Introduction}

To understand the spectacular transport properties of single graphite layers \cite{Nature2007geim,Phys2007geim}
one has to know which kind of scatterers dominates in a given sample.
Indeed, theory predicts the electrical conductivity to be strongly dependent
on the particular type of the scatterers  being present in the system
\cite{cond-mat2007castroneto}.
We demonstrate that the diversity of the conductivity behaviour observed in graphene
\cite{cond-mat2007chen,Nature2007schedin,PRL2007tan,cond-mat2008bolotin,cond-mat2008du}
can be naturally described employing the concentrations of long and
short range scatterers as parameters. According to our microscopic model,
the conductivity measurements upon potassium doping \cite{cond-mat2007chen} obviously suggest a strong domination
of the long range scattering potential at all possible carrier concentrations,
whereas the nitrogen dioxide doping \cite{Nature2007schedin} induces charged
impurities dominating the scattering mechanisms near the conductivity minimum only.
In addition, we shall see that zero chemical doping gives rise to the very sharp minimum conductivity
dip observed in Refs.~\cite{cond-mat2008bolotin,cond-mat2008du}. Our semiclassical approach fully
incorporates the chiral nature of electronic states in graphene which is
crucial for a correct description of transport properties.

\begin{figure}
\includegraphics{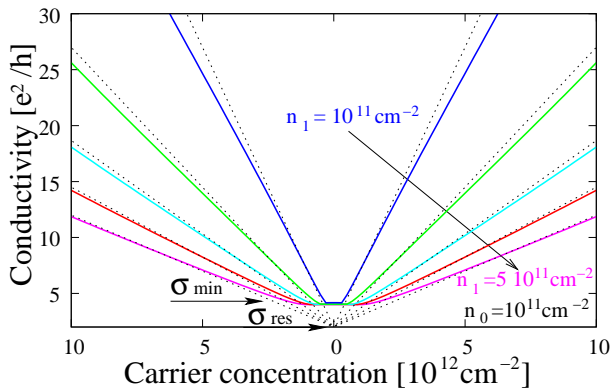}
\caption{Conductivity vs. carrier concentration: Domination of the long range scatterers with
the renormalised concentration $n_1$ [defined below eqs.~(\ref{tau0}--\ref{tau1})]
which increases for each curve from $0.1$ (upper curve)
to $0.5$ [$\times 10^{12}\,\mathrm{cm^{-2}}$] with 
$0.1\cdot 10^{12}\,\,\mathrm{cm^{-2}}$ step. The renormalised concentration of the short range scatterers $n_0$
[defined below eqs.~(\ref{tau0}--\ref{tau1})]
keeps constant at $0.1\cdot 10^{12}\,\mathrm{cm^{-2}}$. Dotted lines are the linear approximations given by 
$\sigma^\mathrm{long}=e\mu^\mathrm{long} n + \sigma^\mathrm{long}_\mathrm{res}$
where $\mu^\mathrm{long}=e/(2hn_1)$ (with $h=2\pi\hbar$ being the Planck constant)
is the carrier mobility obviously governed by the long range scatterers.
The conductivity minimum $\sigma_\mathrm{min}$ is close to $4e^2/h$ whereas
the residual conductivity $\sigma_\mathrm{res}$
is equal to $2e^2/h$ and turns out to be independent of $n_1$.
The sharp peak at $n=0$ due to the term $\propto 2n_1/n$ in eq.~(\ref{maineq})
has been removed as an artifact.
This transport regime corresponds to the graphene samples
strongly doped by potassium \cite{cond-mat2007chen}.}
\label{fig1}
\end{figure}

\begin{figure}
\includegraphics{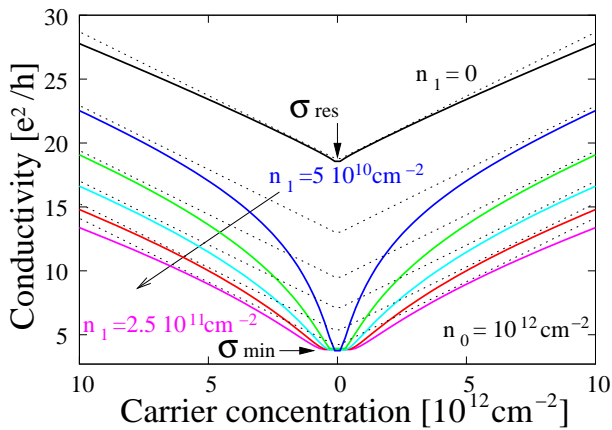}
\caption{Conductivity vs. carrier concentration: 
Both long and short range scattering potentials contribute essentially.
The renormalised concentration of short range scatterers $n_0$
is assumed to be a constant $10^{12}\,\mathrm{cm^{-2}}$, whereas
the concentration of long range scatterers $n_1$ increases for each curve from zero (upper curve)
to $0.25\cdot10^{12}\,\mathrm{cm^{-2}}$] with $0.05\cdot 10^{12}\,\mathrm{cm^{-2}}$ step.
Dotted lines are the linear approximations
given by $\sigma^\mathrm{short}=e\mu^\mathrm{short} n + \sigma^\mathrm{short}_\mathrm{res}$
with the carrier mobility $\mu^\mathrm{short}=e/(hn_0)$ governed by the short range scatterers
and $\sigma^\mathrm{short}_\mathrm{res}$ defined by fit.
Though the conductivity minimum is still near
$4e^2/h$ and does not depend strongly on the impurity
concentration $n_1$, the residual conductivity depends on
both concentrations for short and long range scatterers, see the main text.
This transport regime corresponds to the graphene samples
doped by nitrogen dioxide \cite{Nature2007schedin}.}
\label{fig2}
\end{figure}

There are three most important regimes to consider:
(I) --- {\em Long range scattering potential limit.}
The long range potential of charged impurities strongly dominates the scattering processes and
governs {\em both} the carrier mobility and the conductivity minimum, as shown in fig.~\ref{fig1}
and observed in Ref.~\cite{cond-mat2007chen}.
Here, the width of the minimum conductivity region is rather broad, with a well defined plateau, and 
the conductivity minimum itself is around the universal value $4e^2/h$.
Upon further chemical doping the carrier mobility decreases,
the conductivity as a function of carrier concentration
above its minimum becomes more linear, whereas the residual conductivity defined by
the linear fit keeps constant near $2e^2/h$.
(II) --- {\em Both long and short range scattering potentials are in play.}
At somewhat lower charged impurity concentration, the long range scatterers do not play an essential role
in the conductivity behaviour far away from the conductivity minimum.
Here, the short range scatterers (i. e. nano-sized ripples \cite{Nature2007meyer}
and other imperfections \cite{Philos2007katsnelson}) govern the carrier mobility which is
defined by the linear fit as shown fig.~\ref{fig2} and, thus, turns out to be insensitive to the chemical doping,
as observed in Ref.~\cite{Nature2007schedin}. The minimum conductivity dip is rather sharp, i. e.
no plateau is unambiguously visible. The conductivity minimum is still around $4e^2/h$,
whereas the residual conductivity can acquire a wide range of values depending on the scattering parameters.
(III) --- {\em Short range scattering potential limit.}
If the concentration of the long range scatterers vanishes then
the minimum conductivity dip becomes very sharp as it is shown
in fig.~\ref{fig3} and observed in Refs.~\cite{cond-mat2008bolotin,cond-mat2008du}.
Moreover, at zero concentration of long range scatterers the conductivity minimum becomes
indistinguishable from its residual value
which is not universal and can reach the values up to tens of
$e^2/h$ as observed in Ref.~\cite{PRL2007tan}.
Thus, in this latter case both the minimum conductivity dip and carrier mobility
are governed by the short range scatterers.

To describe these three cases theoretically 
we choose a quasiclassical Boltzmann approach since it allows us to get
transparent analytical formulas for the electrical conductivity in
a broad range of parameters.
The most of the previous attempts 
\cite{PRL2006nomura,JPJ2006ando,PRL2007hwang,PRB2007stauber,PNAS2007adam} based on the Boltzmann equation
discard the chirality of low energy excitations, and we argue
that a proper model must necessary contain {\em both}
peculiar properties of carriers in graphene, namely,
the linear spectrum of low energy excitations {\em and} its chiral nature.
The quasiclassical approach looks at the first sight inapplicable to investigate
the conductivity near its minimum since it is expected to fail at low Fermi energies $E_F$
as soon as $E_F\tau(E_F)$ becomes comparable with $\hbar$.
Note, however, that the relaxation time $\tau$ for short range scatterers
diverges \cite{cond-mat2007castroneto} at $E_F=0$  that leads to
a constant product $E_F\tau(E_F)$ larger than $\hbar$.
Moreover, as we shall see below, for the long range impurity potentials 
the conductivity approaches its minimum away from the Dirac point 
which makes the quasiclassical model applicable again.

We start from the Dirac Hamiltonian $H_0=\hbar v_0 (\sigma_x k_x+\sigma_y k_y)$
describing the low energy excitations in the $\pi$-system of graphene
around the $K$ corner of the first Brillouin zone.
[Here $v_0\approx 10^6 \mathrm{ms^{-1}}$ is the effective ``speed of light'',
$\sigma_{x,y}$ are the Pauli matrices describing the 
pseudo-(or sublattice-)spin, and ${\mathbf k}$ is 
the two-component particle momentum.]
To simplify the model we adopt the valley and spin degeneracy of the electron states,
see Appendix for details.
The eigenstates have the form
\begin{equation}
\label{psi}
\Psi_{\mathbf{k}\pm}(x,y)=\frac{1}{\sqrt{2}}{\mathrm e}^{ik_x x+ik_y y}\left(\begin{array}{c}
1 \\ 
\pm {\mathrm e}^{i\theta}
\end{array} \right),
\end{equation}
where $\tan\theta=k_y/k_x$, and the energy spectrum reads
$E_{k\pm}=\pm \hbar v_0 k$. The spinors (\ref{psi}) contain
momentum-dependent phase factor $e^{i\theta}$ which entangles the momentum
and the sublattice degree of freedom. This chirality of electronic states
plays a crucial role in our description.

To consider both short and long range scatterers on the one hand 
and to avoid having too many fitting parameters on the other, we choose
two opposite types of scattering potential, namely,
$U_0(\mathbf{r})=\lim\limits_{\varrho\rightarrow R} U(\mathbf{r})$
(with $R$ being the radius of the short range scatterers),
and $U_1(\mathbf{r})=\lim\limits_{\varrho\rightarrow \infty} U(\mathbf{r})$,
where $U(\mathbf{r})$ is the generic one given by
$U(\mathbf{r})=(u_0/r){\mathrm e}^{-r/\varrho}$.
Assuming that the short and long range scatterers have
the concentrations $N_0$ and $N_1$ respectively, it is easy
to derive the corresponding relaxation times \cite{cond-mat2007castroneto} given by
\begin{eqnarray}
\label{tau0} &&
\tau_0=\frac{1}{4\pi R^2n_0 v_0 k}\frac{2R^4k^4}{1+2R^2k^2-\sqrt{1+4R^2k^2}},\\
\label{tau1} &&
\tau_1=\frac{k}{4\pi n_1 v_0},
\end{eqnarray}
where $n_{0,1}=\pi N_{0,1}(2u_0/\hbar v_0)^2$
are the renormalised concentrations being only the two fitting parameters.
The characteristic size $R$ of ripples and other imperfections
is assumed to be near a few $\mathrm{nm}$.

To incorporate the chiral nature of Dirac fermions 
into the stationary quasiclassical kinetic equation
linear in the homogeneous electric field ${\mathbf E}$
we have to start from its general form
\begin{equation}
\label{keq}
\frac{1}{\hbar}\left(e{\mathbf E} \frac{\partial \hat{f}}{\partial{\mathbf k}} + i\left[H_0,\hat{f}({\mathbf k})\right]
\right)
=\mathrm{I}[\hat{f}],
\end{equation}
where $\hat{f}$ is the density matrix,
$\mathrm{I}[\hat{f}]$ is the collision integral.
In eq.~(\ref{keq}), the sublattice degree of freedoom is assumed to be
a quantum number, although the particle momentum ${\mathbf k}$ is still considered quasiclassically.
To deal with such an equation it is natural
to rewrite it in the basis of the eigenvectors (\ref{psi}).
The distribution function $f$ represents then
a $2\times 2$ matrix, and its off-diagonal elements
do not drop out thanks to the chiral nature of the eigenstates (\ref{psi}).
This important modification of the Boltzmann equation allows us to deal with
the essentially quantum effect suggesting that a particle
could not only be in one of the states
$\Psi_{{\mathbf k}+}$ or $\Psi_{{\mathbf k}-}$ but in an arbitrary superposition of them.
Alternatively one might think about the
off-diagonal elements as a manifestation of the {\em pseudospin}
precession \cite{Phys2007geim} associated with the chirality index $\kappa$.

To solve the kinetic equation with respect to $f$
we assume the linear response regime (small ${\mathbf E}$).
All technical details concerning the kinetic equation,
its solution, and derivation of conductivity expressions
are given in the Appendix section.
To write down the final zero-temperature
conductivity formula it is convenient to introduce
the following parameters: $\gamma_0=\pi R^2 n_0$
and $\gamma_n=\pi R^2 n$, where $n=k_F^2/\pi$ is the carrier concentration
with $k_F=E_F/\hbar v_0$ being the Fermi wave vector.
Then, the conductivity reads
\begin{equation}
\sigma=\frac{e^2}{h}\left\{
\frac{n\gamma_n}{n_0\left(1+2\gamma_n-\sqrt{1+4\gamma_n}\right)+2n_1\gamma_n} + \frac{2n_1}{n}\right.
$$
$$
\left. +4\gamma_0\left[\frac{3}{8}\frac{1}{\gamma_n^2}+\frac{1}{2\gamma_n}+\sqrt{1+4\gamma_n}\left(
\frac{1}{4\gamma_n}-\frac{3}{8}\frac{1}{\gamma_n^2}\right) \right.\right.
$$
$$
\left.\left. - \ln\left(\frac{1}{2\sqrt{\gamma_n}}+\sqrt{1+\frac{1}{4\gamma_n}}\right) \right]\right\}+
\sigma^\mathrm{long}_\mathrm{res}.
\label{maineq}
\end{equation}
The sharp peak at $n=0$ in eq.~(\ref{maineq}) due to the term $\propto 2n_1/n$
should be seen as an artifact since our quasiclassical approach certainly fails in this limit.
According to the experimental data there is normally a plateau instead of that peak.
We define the residual conductivity $\sigma^\mathrm{long}_\mathrm{res}$ from
the linear fit for the conductivity measurements in the limit 
of long range scatterers domination,
as explained in Ref.~\cite{cond-mat2007chen}.
As follows from the measurements \cite{cond-mat2007chen}, $\sigma^\mathrm{long}_\mathrm{res}$ turns out to be 
a constant nearly equal to $2e^2/h$.
The residual conductivity $\sigma^\mathrm{long}_\mathrm{res}$
is the only parameter that could not be derived within
our quasiclassical approach because of the obvious
restriction discussed before.
But it contributes to $\sigma(n)$ in a very simple way:
$\sigma^\mathrm{long}_\mathrm{res}=2e^2/h$ just shifts all
the $\sigma(n)$ curves into the region of higher conductivities
and does not have an influence on the quasiclassical physics
above $\sigma^\mathrm{long}_\mathrm{res}$.
Note, however, that the residual conductivity
is not a constant as soon as the long range scatterers do not dominate any more.
Such residual conductivity
$\sigma^\mathrm{short}_\mathrm{res} > \sigma^\mathrm{long}_\mathrm{res}$
can be derived within our approach since it remains
valid for short range scatterers even at zero carrier concentration.
Note, that eq.~(\ref{maineq}) is true for arbitrary
values of $n_{0,1}$ and $R$, thus,
we can utilise this formula to distinguish the transport regimes
with the domination of short or long range scatterers.
To proceed we set $R=1\,\mathrm{nm}$, thus
$\gamma_n \ll 1$ for all reasonable carrier concentrations,
and, as consequence, $\gamma_0 \ll 1$ as well.

\section{(I) --- Long range scattering potential limit}

Long range scatterers dominate if $n_1 \geq n_0$. Here,
eq.~(\ref{maineq}) can be expanded in terms of $\gamma_n$
and $\gamma_n n_0/n_1$ resulting in the following expression
\begin{equation}
\label{long}
\sigma^\mathrm{long}=\frac{e^2}{h}\left[
\frac{n}{2n_1} + \frac{2n_1}{n}+
\gamma_0 \left(
5+2\ln\gamma_n - \frac{n^2}{2n_1^2} \right)\right]
+ \sigma^\mathrm{long}_\mathrm{res},
\end{equation}
\begin{equation}
\label{long0}
\sigma^\mathrm{long}_{n_0=0}=\frac{e^2}{h}\left(
\frac{n}{2n_1} + \frac{2n_1}{n}\right) + \sigma^\mathrm{long}_\mathrm{res},\quad n_0=0.
\end{equation}
The function $\sigma^\mathrm{long}(n)$ has a minimum
near $n_\mathrm{min}= 2n_1$, $n_1\neq 0$.
At this concentration the term proportional to $\gamma_0$
is much smaller then the leading one,
and the conductivity reaches its minimum
close to $\sigma^\mathrm{long}_\mathrm{min}=\sigma^\mathrm{long}_\mathrm{res}+2e^2/h$
as it is seen in fig.~\ref{fig1}.
Since $\sigma^\mathrm{long}_\mathrm{res}=2e^2/h$ 
the conductivity minimum indeed acquires the typical value
$4e^2/h$ in quite perfect samples
(i. e. without ripples, neutral impurities etc.)
when $\gamma_0\rightarrow 0$.
The corrections of the order of $o(\gamma_0)$ and $o(\gamma_0\ln\gamma_n)$
give all together a small negative contribution to the
conductivity minimum.
Most importantly, the conductivity minimum does not
depend on $n_1$, i. e. it should be insensitive to doping, whereas
the minimum carrier concentration $n_\mathrm{min}=2n_1$
increases linearly with chemical doping
and in that way gives rise to the width
of the minimum conductivity plateau,
as observed in \cite{cond-mat2007chen}.
Above $n=n_\mathrm{min}$ (i. e. $n\gg 2n_1$)
one can linearise eq.~(\ref{long})
and obtain $\sigma^\mathrm{long}=e\mu^\mathrm{long} n + \sigma^\mathrm{long}_\mathrm{res}$
with $\mu^\mathrm{long}=e/(2hn_1)$ being the carrier mobility.
The deviations from linear dependency $\sigma^\mathrm{long}(n)$
are described by the term $\propto -n^2/2n_1^2$ in eq.~(\ref{long})
and can be seen in fig.~\ref{fig1} as well.
All these peculiar features indicate the domination of long range scatterers
that was observed in \cite{cond-mat2007chen}.

\section{(II) --- Both long and short range scattering potentials are in play}

At lower charged impurity concentrations
the minimum conductivity dip becomes sharper,
the plateau vanishes, and the carrier mobility
is governed by the short range scatterers, see fig.~\ref{fig2}.
For large enough carrier concentrations so that $n_0\gamma_n\gg n_1$
the conductivity can be written as
\begin{equation}
\label{short}
\sigma^\mathrm{short}=\frac{e^2}{h}
\left[\frac{1}{2\gamma_0}\left(1-\frac{n_1}{n_0}\frac{1}{\gamma_n}\right) +\frac{n}{n_0}\right]
+ \sigma^\mathrm{long}_\mathrm{res}.
\end{equation}
We emphasise that eq.~(\ref{short}) is just an estimate
for the conductivity far away from its minimum, i. e.
for the carrier concentrations $n\gg n_1/\gamma_0$.
Eq.~(\ref{short}) allows us to define the carrier mobility as
$\mu^\mathrm{short}=e/(hn_0)$ and residual conductivity
$\sigma^\mathrm{short}_\mathrm{res}=
\frac{e^2}{h}\frac{1}{2\gamma_0}\left(1-\frac{n_1}{n_0}\frac{1}{\gamma_N}\right)+\sigma^\mathrm{long}_\mathrm{res}$
with $\gamma_N$ being $\gamma_n$ taken at $n=N$ large enough
to satisfy $N\gg n_1/\gamma_0$.
Though the estimation for $\sigma^\mathrm{short}_\mathrm{res}$ is very rough 
[since eq.~(\ref{short})
is an estimate as well], it shows that the
residual conductivity is not a constant anymore
as long as the short range scatterers dominate.
Thus, the residual conductivity depends on both short and long range scatterers,
whereas the conductivity minimum is governed by the long range scatterers alone
and remains near the typical value $4e^2/h$, as one can observe in fig.~\ref{fig2}.
The latter is due to the fact that $\tau_1\ll\tau_0$ at vanishing carrier density
even if $n_0\gg n_1$. On the other hand, the carrier mobility $\mu^\mathrm{short}$ depends on $n_0$ and, thus,
is unaffected by the chemical doping. Such behaviour indicates the short range scatterers domination
in the conductivity measurements upon nitrogen dioxide doping \cite{Nature2007schedin}.

\section{(III) --- Short range scattering potential limit}

\begin{figure}
\includegraphics{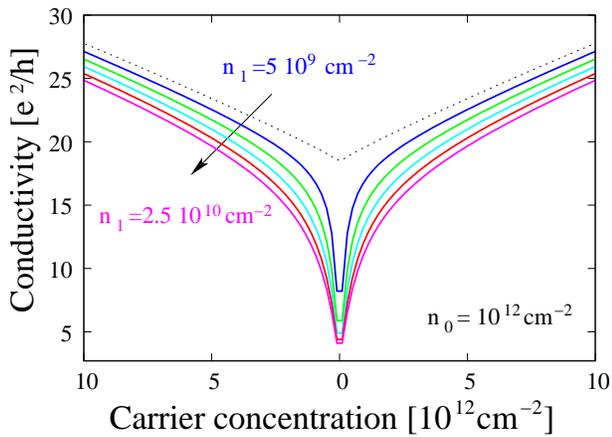} 
\caption{Conductivity vs. carrier concentration:
Domination of the short range scatterers with
the renormalised concentration $n_0=10^{12}\,\mathrm{cm^{-2}}$.
The renormalised concentration of long range scatterers $n_1$ increases for each curve from
$5\cdot10^{9}\,\mathrm{cm^{-2}}$  (upper curve)
to $2.5\cdot10^{10}\,\mathrm{cm^{-2}}$ with $5\cdot 10^{9}\,\mathrm{cm^{-2}}$ step.
The dotted line corresponds to the case $n_1=0$.
The conductivity dip is very sharp and does not contain a plateau.
According to the very recent report \cite{cond-mat2008bolotin},
the conductivity of suspended graphene after annealing indeed demonstrates
very sharp dip at zero gate voltages.
Comparison of the conductivity measurements for suspended and non-suspended graphene
\cite{cond-mat2008du} is also consistent with our predictions.}
\label{fig3}
\end{figure}

It is instructive to consider the special case $n_1 = 0$
in detail. Here the term $\propto 2n_1/n$ does not dominate over
the logarithmic one $\propto \ln\gamma_n$ at $n\rightarrow 0$,
and, therefore, the latter must be retained. The conductivity then reads
\begin{equation}
\label{short0}
\sigma^\mathrm{short}_{n_1=0}=\frac{e^2}{h}
\left(\frac{1}{2\gamma_0}+ \frac{n}{n_0} + 2\gamma_0\ln\gamma_n \right)
+ \sigma^\mathrm{long}_\mathrm{res}.
\end{equation}
Interrestingly, the conductivity minimum
is uncertain, but since $R$ is small ($\gamma_{0,n}\ll 1$)
the conductivity at $n\rightarrow 0$ is indeed indistinguishable 
from its residual value
$\sigma^\mathrm{short}_\mathrm{res}(n_1=0)=\frac{e^2}{h}\frac{1}{2\gamma_0}+\sigma^\mathrm{long}_\mathrm{res}$.
This case is depicted in fig.~\ref{fig3} and corresponds to the chemically undoped \cite{PRL2007tan}
and suspended \cite{cond-mat2008bolotin,cond-mat2008du} graphene samples
where the conductivity demonstrates rather sharp dip
with nonuniversal minimum value strongly dependent on scattering parameters.
It is noteworthy that similar logarithmic corrections
to the conductivity minimum have been found
in presence of electron-electron interactions \cite{PRL2007mishchenko,PRL2008herbut}
and scattering on vacancies, cracks, or boundaries \cite{PRB2007stauber}.

Given the above theoretical results, let us now analyse recent
experiments on graphene samples upon chemical doping
\cite{Nature2007schedin,cond-mat2007chen}.
We first discuss the case of potassium doping \cite{cond-mat2007chen}.
Here, the width of the minimum conductivity region broadens 
(i. e. $n_\mathrm{min}$ increases) whereas both the minimum conductivity
and mobility decrease, at least initially.
At certain stage, however, the conductivity minimum
becomes insensitive to further doping.
These features are clearly understood within our model.
Initially, the sample is very weakly doped, and
the minimum conductivity is governed by the short range scatterers
rather than charged impurities in accordance with eq.~(\ref{short0}).
At larger potassium concentrations
the long range scatterers start to dominate,
the minimum concentration $n_\mathrm{min}\propto n_1$
increases whereas the conductivity minimum keeps constant
despite further decreasing the mobility.
This limit corresponds to  eq.~(\ref{long0}).
In Ref.~\cite{Nature2007schedin}, graphene is dosed by nitrogen dioxide.
Here, the width of the conductivity minimum region broadens
similar to the previous case but it remains much more narrow
than upon potassium doping.
Moreover, the carrier mobility and conductivity minimum
are both insensitive to doping.
According to eq.~(\ref{short}) such data suggest strong domination 
of the short range scatterers,
at least at higher carrier concentrations $n \gg n_1/\gamma_0$.
The reason of such a strong suppression of induced charges
might be the strong screening of Coulomb interactions
in the samples examined. Such precise description of charged
impurities is certainly beyond the scope of this report, where $U_1$ is not screened at all.
We can predict, however, that the minimum conductivity phenomenon
takes place even in free hanging graphene sheets, without
any influence of the silicon dioxide substrate considered in \cite{PNAS2007adam}.

Moreover, our model offers a solution to
the urgent minimum conductivity problem.
Previous measurements appear to be contradictory in so far as
quite a representative
group of graphene samples \cite{Nature2007geim}
exhibits the conductivity minimum
clustering around $4e^2/h$,
whereas many other samples \cite{PRL2007tan} demonstrate
very wide range of the minimum conductivity values, away from $4e^2/h$.
Now it is clear that the universal conductivity minimum
in the samples from \cite{Nature2007geim} is 
due to the long range scatterers (e. g. charged impurities) with
a little influence of ripples and other short range imperfections.
This results in the conductivity dip broaden enough
to resolve its bottom at $\sigma \sim 4e^2/h$.
(The carrier mobility, however, is not necessarily governed
by charged impurities.)
In contrast, the conductivity minimum far away
from $4e^2/h$ betrays the lack of the charged impurities
in a given sample that makes the conductivity dip sharper and 
its bottom difficult to resolve, as it is seen from eq.~(\ref{short0}).
Thus, the minimum conductivity becomes indistinguishable from
its residual value which, in general, is sensitive to the concentrations of both long and short range scatterers.
It is therefore not surprising that the conductivity minimum
changes from sample to sample in this case. In that way our model is able to explain
the wide diversity of the minimum conductivity values measured in single layer graphene.

As an improtant further conclusion we would like to emphasise that
it is the chiral nature of low energy excitations,
rather than its linear spectrum, that is responsible for
the minimum conductivity phenomenon in graphene.
It is especially obvious for quite perfect
graphene samples where long range scatterers (i. e. charged impurities)
dominate the scattering processes: The conductivity would be given
just by the first term in eq.~(\ref{long0}), and the conductivity minimum would certainly vanish
as long as the chiral nature were discarded.
Such a conclusion is consistent with the theoretical investigation
of ballistic graphene \cite{EPJB2006katsnelson} and strongly supported by the very experimental fact
that bilayer graphene, despite obviously non-linear dispersion law,
also exhibits a minimum conductivity of the same order as single layer samples \cite{Nature2007geim}.
The conductivity measurements for chemically doped bilayer samples
are not accessible yet, they would certainly be
an {\em experimentum crucis} on which the mechanism is 
responsible for the conductivity minimum in graphene.

\section{Appendix --- Solution of the kinetic equation}

The distribution function $f({\mathbf k})$ for carriers in presence of the electric field $\mathbf{E}$ can be written as
the sum of an equilibrium contribution $f^0(E_{k \kappa})$ (which is just the Fermi function)
and a nonequilibrium part $f^1({\mathbf k})$ which represents a $2\times 2$ matrix.
To find $f^1(\mathbf{k})$ one has to solve eq.~(\ref{keq}) written in the helicity basis (\ref{psi}).
To do that one has to bear in mind that the spinors are ${\mathbf k}$-dependent (via the phase $i\theta$),
thus, the kinetic equation differs from its standart analogue and reads
\begin{eqnarray}
\nonumber && 
\mathrm{I}[f]=
\frac{i}{\hbar}\left(
\begin{array}{cc}
0 & f_{12}\left(E_{k+}-E_{k-}\right) \\  f_{21}\left(E_{k-}-E_{k+}\right) & 0
\end{array}\right) \\
\nonumber &&
+e{\mathbf E}\left(
\begin{array}{cc}
 -\mathbf{v}_{11} \left[-\frac{\partial f^0(E_{k +})}{\partial E_{k +} }\right] &
\frac{\mathbf{v}_{12}}{2E_{k+}}\left( f^0_{E_{k -}} - f^0_{E_{k +}} \right) \\ 
\frac{\mathbf{v}_{21}}{2E_{k-}}\left( f^0_{E_{k +}} - f^0_{E_{k -}} \right)
& -\mathbf{v}_{22} \left[-\frac{\partial f^0(E_{k -})}{\partial E_{k -} }\right]
\end{array}\right)\\
\label{master1}
\end{eqnarray}
with $\mathbf{v}_{ij}$ being the velocity matrix elements given by
\begin{equation}
\label{v}
\frac{{\mathbf v}}{v_0} ={\mathbf e_x}\left(
\begin{array}{cc}
\cos\theta  & -i\sin\theta \\ 
i\sin\theta & -\cos\theta
\end{array}\right) + {\mathbf e_y} \left(
\begin{array}{cc}
\sin\theta & i\cos\theta \\
-i\cos\theta & -\sin\theta
\end{array}\right).
\end{equation}
One might ask why spin and valley degrees of freedom
are discarded in eq.~(\ref{master1}).
The reason is that the total Hamiltonian $H$
(which includes, in addition to the sublattice index $\kappa$,
the spin $s$ and valley $\nu$ indices) does not
contain the crossing (off-diagonal) terms $H^{ss'}$ and $H^{\nu\nu'}$,
and, therefore spin and valley coherence does not manifest itself in the conductivity.
In contrast, the sublattice degree of freedom couples electron
and hole states via the off-diagonal terms $H_0^{\kappa\kappa'}$
in the Hamiltonian $H_0$ that leads to the non-diagonal velocity operator and electron-hole
coherent contributions in the conductivity.

To find the collision integral in eq.~(\ref{master1}) we start from
the commutator $-\frac{i}{\hbar}[U,\hat{f}]$, where 
the density matrix is substituted by the following expression describing its time
evolution after each scattering event
\begin{equation}
 \hat{f}\rightarrow\hat{f}(t=0)+
\frac{i}{\hbar}\int\limits_0^\infty dt {\mathrm e}^{-\frac{i}{\hbar}H_0 t}
[\hat{f}(t=0),U]{\mathrm e}^{\frac{i}{\hbar}H_0 t}.
\end{equation}
The integrals over $t$ can be taken introducing a small
parameter in the time dependent exponent
\begin{equation}
\label{intt}
\int\limits_0^\infty dt {\mathrm e}^{\frac{i}{\hbar}(E_{{\mathbf k'}\kappa'}-E_{{\mathbf k}\kappa})t}
=P \frac{i\hbar}{E_{{\mathbf k'}\kappa'}-E_{{\mathbf k}\kappa}}+\pi\hbar\delta(E_{{\mathbf k'}\kappa'}-E_{{\mathbf k}\kappa}),
\end{equation}
where $P$ denotes the principle value.
Then the collision integral reads
\begin{eqnarray}
\nonumber
&&
\mathrm{I}[f]_{\kappa\kappa'}=
\int\frac{d^2{\mathbf k'}}{(2\pi)^2}\sum\limits_{\kappa_1,\kappa'_1}
\{[\delta(E_{k' \kappa_1}-E_{k \kappa}) \\
\nonumber &&
+\delta(E_{k'\kappa'_1}-E_{k\kappa'})]
K^{\kappa\kappa'}_{\kappa_1 \kappa'_1}f_{\kappa_1 \kappa'_1}(k')
-\delta(E_{k\kappa_1}-E_{k' \kappa'_1}) \\
&&
\times [K^{\kappa\kappa_1}_{\kappa'_1 \kappa'_1}f_{\kappa_1 \kappa'}(k)+f_{\kappa \kappa_1}(k)
K^{\kappa_1 \kappa'}_{\kappa'_1 \kappa'_1}]\},
\label{I}
\end{eqnarray}
with $K^{\kappa\kappa'}_{\kappa_1\kappa'_1}$ being
\begin{eqnarray}
\nonumber &&
K^{\kappa\kappa'}_{\kappa_1\kappa'_1}
=\frac{\pi}{\hbar}\sum\limits_{i=0,1}
\left<\Psi_{\kappa k}\mid U_i\mid \Psi_{\kappa_1 k'} \right>
\left<\Psi_{\kappa' k}\mid U_i\mid \Psi_{\kappa'_1 k'} \right>^*.
\end{eqnarray}
This form of the collision integral
describes elastic scattering of chiral particles within Born approximation
and is routinely applied
to kinetics of chiral electrons with spin-orbit interactions,
see e.g. \cite{JETP1984dyakonov,PRL2006khaetskii}.
It is important that the intrinsicly inelastic terms [i. e. the principle value in eq.~(\ref{intt})]
are discarded in the final expression for $I[f]$.
These terms describe creation of electron-hole
pairs and have been dubbed in \cite{PRB2007auslender} as {\em zitterbewegung} contributions.
The latter does not change results qualitatively but
essentially complicates the analytics as one can see
comparing Refs. \cite{PRB2007auslender} and \cite{PRL2007trushin}.
To reach a compromise between transparency and completeness of our model
we deal only with the leading (i. e. $\delta$-functional) terms in the collision integral.

The solution of eq.~(\ref{master1}) can be written down as 
$f({\mathbf k})=f^0(E_{k \kappa})+f^1({\mathbf k})$, where
\begin{eqnarray}
\label{f11} \nonumber && 
f_{11}^1=q\mathbf{E}\mathbf{v}_{11} \tau(k)\left\{\left(1+\frac{1}{2\alpha}\right)
\left[-\frac{\partial f^0(E_{k+})}{\partial E_{k+} }\right]\right.\\
&&
\left.
+ \frac{1}{2\alpha} \left[-\frac{\partial f^0(E_{k-})}{\partial E_{k-} }\right]
+\frac{1}{2\alpha E_{k+}}\left( f^0_{E_{k +}} - f^0_{E_{k -}} \right)\right\} \\
\label{f12} \nonumber &&
f_{12}^1=\frac{q\mathbf{E}\mathbf{v}_{12} \tau(k)\left(\frac{1}{2}+\frac{1}{2\alpha}\right)}{1+2iE_{k+}\tau(k)/\hbar}
\left\{\frac{1}{E_{k+}}\left( f^0_{E_{k +}} - f^0_{E_{k -}} \right)
\right.\\
&&
\left.
+\left[-\frac{\partial f^0(E_{k+})}{\partial E_{k+} }-\frac{\partial f^0(E_{k-})}{\partial E_{k-} }\right]\right\},
\end{eqnarray}
and $f_{22}^1$, $f_{21}^1$ can be obtained from eqs.~(\ref{f11}--\ref{f12})
just exchanging the indices belong to $E_k$ and $\mathbf{v}$ accordingly.
Here, $\tau^{-1}=\tau_0^{-1}+\tau_1^{-1}$ is the total momentum
relaxation time, and $\alpha(k)=4E_{k+}^2\tau^2/\hbar^2$
is the electron-hole incoherence parameter.
The off-diagonal elements $f_{12}^1$ and $f_{21}^1$ do not contribute to the current
if and only if they become real at $\alpha\gg 1$.

The electrical current
$\mathbf{j}=e\int\frac{d^2 k}{(\pi)^2}\mathrm{Tr}\left[\mathbf{v}(k)f^1(\mathbf{k})\right]$
is easy to derive at low temperatures ($T \ll E_F$), and the conductivity
can be written as a sum $\sigma=\mathbf{j}/\mathbf{E}+\sigma^\mathrm{long}_\mathrm{res}$
with a constant $\sigma^\mathrm{long}_\mathrm{res}$ being the residual
conductivity in the limit of the long range scatterers domination
defined by the linear fit from the experimental data \cite{cond-mat2007chen}.
The formula (\ref{maineq}) follows directly from eqs.~(\ref{f11}--\ref{f12})
and is exact in this sense. It is only essential to integrate over ${\mathbf k}$
{\em before} the subsequent approximations relying on a small $R$;
we would obtain the logarithmic divergence in the conductivity integral otherwise 
\cite{PRB2007auslender,PRL2007trushin}.

\acknowledgments

This work was financially supported by SFB 689.
We thank Mikhail Katsnelson and Tobias Stauber for stimulating discussions.

\end{document}